\documentclass[preprint,showpacs,preprintnumbers,amsmath,amssymb]{revtex4}

\usepackage{amsmath}
\usepackage{amssymb}
\usepackage{amsthm}
\usepackage{amsfonts}
\usepackage{graphicx}
\usepackage{dcolumn}
\usepackage{bm}
\usepackage{epsfig}

\begin{document}
\title{The asymptotic Bethe ansatz solution for
one-dimensional $SU(2)$ spinor bosons with finite range Gaussian
interactions}

\author{J. Y. Lee$^{1}$,
 X. W. Guan$^{1}$, A. del Campo$^{2}$
 and M. T. Batchelor$^{1,3}$}

 \affiliation{$^{1}$ Department of Theoretical Physics,
 Research School of Physics and Engineering,
 Australian National University, Canberra ACT 0200, Australia}

 \affiliation{$^{2}$ Institut f\"{u}r Theoretische Physik, Leibniz
 Universit\"{a}t Hannover, Hannover, Germany}

 \affiliation{$^{3}$ Mathematical Sciences Institute,
 Australian National University, Canberra ACT 0200, Australia}

\date{\today}

\begin{abstract}
We propose a one-dimensional model of spinor bosons with $SU(2)$
symmetry and a two-body finite range Gaussian interaction potential.
We show that the model is exactly solvable when the width of the
interaction potential is much smaller compared to the inter-particle
separation. This model is then solved via the asymptotic Bethe
ansatz technique. The ferromagnetic ground state energy and chemical
potential are derived analytically. We also investigate the effects
of a finite range potential on the density profiles through local
density approximation. Finite range potentials are more likely to
lead to quasi Bose-Einstein condensation than zero range potentials.
\end{abstract}



\maketitle

\section{Introduction}
Integrable one-dimensional (1D) models of interacting bosons and
fermions with $\delta$-function interaction
\cite{Lieb1963,Yang1967,Gaudin1967} have had a tremendous impact on
quantum statistical mechanics. In particular, recent breakthrough
experiments on trapped ultracold bosons and fermions atoms confined
to 1D have provided a better understanding of quantum statistical
effects and strongly correlated phenomena in quantum many-body
systems. These models contain two-body zero range potentials which
allows the wavefunctions to be written as a superposition of plane
waves by means of Bethe's hypothesis \cite{Bethe1931}. This
assumption is true based on the fact that every particle can move
freely without feeling the presence of others when no collision
takes place.

However, Calogero \cite{Calogero1969} showed that certain models
with long range potentials can also be solved exactly, though not
using Bethe's hypothesis. He first solved the three-body problem
with a harmonic potential and a $g/r^{2}$ potential, and then
generalized it to the $N$-body problem to obtain the exact
expression for the ground state energy and a class of excited
states. Sutherland \cite{Sutherland1971} then derived the exact
solutions for the ground state energy, pair correlation function,
low-lying excitations and thermodynamics of the model with $g/r^{2}$
potential for both fermions and bosons in the thermodynamic limit by
employing the \emph{asymptotic Bethe ansatz} (ABA) which uses
Bethe's hypothesis in the asymptotic limit. Since then, many models
with non-local interaction were solved exactly through the ABA
method. Among them are the isotropic Heisenberg antiferromagnetic
chain \cite{Haldane1988}, the quantum lattice model with inverse
$\sinh$ square potential \cite{Sutherland1994}, the $t-J$ model with
long range interaction \cite{Kawakami1992}, the nonliner
Schr\"{o}dinger model \cite{Kundu1993} and so on.

The main idea of the ABA is that one restricts oneself to the
asymptotic region where the particles are considered to be
sufficiently far apart, such that their influence on neighboring
particles is negligible \cite{Sutherland}. Then one has to show by
some unspecified method that the system is integrable, i.e., that it
has a complete set of independent integrals of motion. For example,
various authors \cite{Moser1975} have shown that for $g/r^{2}$
potentials, one can find $N$ integrals of motion for the $N$
particle system. Once this is done, one can then conclude that the
wavefunction is non-diffractive and thus asymptotically given by the
BA. Since the exact scattering data is known, one can then obtain
the exact thermodynamics of the system \cite{Sutherland1993}. It
should be pointed out that a common misconception is that the ABA is
only a low-density approximation, i.e., $N/L\rightarrow 0$. This is
not true and in fact it gives the exact thermodynamics for systems
with finite density in the thermodynamic limit (see
\cite{Sutherland} for explanations). When using the ABA, the
low-density limit $N/L\rightarrow 0$ is only reached when the width
of the interaction potential between neighboring particles become
large. However, for the purpose of this investigation, we 
restrict ourselves to a finite density system where the width of the
interaction potential between particles is small. A physical example
of systems with such properties are dilute gases, whose
inter-particle interactions are almost local.

In this paper, we investigate the ground state of two-component
spinor bosons with finite range Gaussian interactions in 1D. The
interaction potential for this system can be expressed in terms of
the sum of even powered derivatives of a $\delta$-function. It gives
rise to certain nonlinear behaviour not observed in systems
with spin-independent potentials \cite{Batchelor2008}. This kind of
velocity- or state-dependent potential leads to more versatility in
studying spin waves, ferromagnetic behaviour  and the relation
between superfluidity and magnetism in low-dimensional many-body
systems, as shown in Ref.~\cite{Wicke2010} for two-component
$^{87}$Rb atoms on a quantum chip. By using a state-dependent
dressed potential, spin degrees of freedom in two-component spinor
bosons are tunable. This technique for controlling
non-equilibrium spin motion allows one to study quantum coherence in
interacting quantum systems, and to experimentally explore
predictions of the thermodynamic Bethe Ansatz (TBA) in a system of
two-component spinor bosons.

We first introduce the model in Section \ref{Model}. In Section
\ref{Integrability}, we show that the Hamiltonian for this model is
integrable. In Section \ref{GroundState}, we derive the distribution
functions for the charge and spin degrees of freedom from the ABA
equations. The ground state energy and thermodynamics are evaluated 
in Section \ref{Thermodynamics} 
in the limits where the interaction strength between particles is
large and the width of the interaction potential is small. In Section \ref{Local}, we apply the
local density approximation to obtain the density profiles for this
model. And finally in Section \ref{Conc}, we conclude with a summary
of our results.

\section{The Model}
\label{Model} Let us consider $N$ bosons with $SU(2)$ symmetry
confined to a 1D wire of length $L$ with periodic boundary
conditions. Here we denote the internal hyperfine spin states as
$|\uparrow\rangle$ and $|\downarrow\rangle$. The interaction
potential between adjacent particles is given by a generic
non-negative function $v(x_{j}-x_{l})$ that is even in the
inter-particle separation, i.e., $v(x)=v(-x)$ and vanishes at large
enough distances, i.e., $\lim_{x\rightarrow\infty}v(x)=0$. For such
a system, the first quantized Hamiltonian is given by
\begin{equation}
\mathcal{H}=-\frac{\hbar^{2}}{2m}\sum_{j=1}^{N}\frac{\partial^{2}}{\partial
x_{j}^{2}}+2c\sum_{j<l}v(x_{j}-x_{l})-\frac{H}{2}(N_{\uparrow}-N_{\downarrow}),\label{Ham1}
\end{equation}
where $m$ is the mass of each boson and $c$ characterizes the
interaction strength which is the same for all possible collisions,
i.e., between two $|\uparrow\rangle$ bosons, two
$|\downarrow\rangle$ bosons, or one $|\uparrow\rangle$ and one
$|\downarrow\rangle$ boson. The interactions are repulsive when
$c>0$ and attractive when $c<0$. The external magnetic field is
represented by $H$, and the total particle number is given by
$N=N_{\uparrow}+N_{\downarrow}$. For the rest of this paper we use
the dimensionless units of $\hbar=2m=1$ for convenience. 
These units are also used in all figures. 

In the case when
\begin{equation}
v(x)=\frac{1}{\sqrt{2\pi\alpha^{2}}}\exp\left(-\frac{x^{2}}{2\alpha^{2}}\right),
\label{gaussPotential}
\end{equation}
the model can be exactly solved in the region $x_{1}\ll x_{2}\ll
\ldots\ll x_{N}$ where the width of the Gaussian potential $\alpha$
is small relative to the inter-particle separation, i.e.,
$|x_{i+1}-x_{i}|\gg\alpha$ or $(N/L)\alpha\ll 1$ for every $i<N$. In
this limit, all particles scatter non-diffractively. This implies
that the asymptotic wavefunction can be written as a sum of $N!$
terms corresponding to the permutations $P$ of the set of asymptotic
momenta $\{k_{i}\}$. Explicitly, the wavefunction can be expressed
in Bethe ansatz form as
\begin{equation}
\psi(\mathbf{x})=\sum_{P}A(P)\exp\left(\mathrm{i}\sum_{j=1}^{N}k_{P_{j}}x_{j}\right).
\end{equation}

The argument that supports non-diffractive scattering is as follows.
Consider the two-body problem $N=2$ where the particles are far
apart, i.e., $x_{1}\ll x_{2}$. Since $|x_{2}-x_{1}|\gg\alpha$, the
particles behave as free particles, therefore the wavefunction is a
product of plane waves with total momentum and energy given by
\begin{equation}
P=k_{1}+k_{2}, \qquad E=k_{1}^{2}+k_{2}^{2}.
\end{equation}
Through the scattering process, the total momentum and energy have
to be conserved. This yields a new set of momenta which is either
$(k_{1}',k_{2}')=(k_{1},k_{2})$ or $(k_{1}',k_{2}')=(k_{2},k_{1})$.

For the $N$-body problem, we can think of it as a succession of two
particles colliding and then scattering to the asymptotic region as
free particles, where each two-body collision gives rise to a
permutation of the momenta. A product of transpositions acting on
the permutation $P$ leads to another permutation $P'$. Hence, the
scattering is non-diffractive for any number of particles. When
$\alpha\rightarrow 0$ in the fully polarized case, $v(x)\rightarrow\delta(x)$ which allows us to
recover the Lieb-Liniger interacting spinless Bose gas
\cite{Lieb1963}.

\section{Integrability of the Hamiltonian}
\label{Integrability} We know that in the limit $\alpha\rightarrow
0$, the Gaussian function tends to the $\delta$-function. The
$\delta$-function is not a function in the classical sense, and
should be treated as a generalized function \cite{Gelfand1964}
instead. Notice that if the potential $v(x)$ is an even function,
its Fourier transform
$\widehat{v}(k)=\int_{-\infty}^{\infty}v(x)e^{\mathrm{i}kx}dx$ is
also an even function, i.e., $\widehat{v}(k)=\widehat{v}(-k)$. This
implies that the Taylor expansion of $\widehat{v}(k)$ in the
neighborhood of $k=0$ only consists of even powers of $k$ as given
by
\begin{equation}
\widehat{v}(k)=\sum_{n=0}^{\infty}b_{n}k^{2n}. \label{FTv}
\end{equation}
Assuming that the potential meets such restrictions, we can take the
inverse Fourier transform to obtain the potential in position space
as
\begin{eqnarray}
\nonumber v(x) &=&
\frac{1}{2\pi}\sum_{n=0}^{\infty}\int_{-\infty}^{\infty}b_{n}k^{2n}e^{-\mathrm{i}kx}dk
\\ &\equiv& \sum_{n=0}^{\infty}a_{n}\delta^{(2n)}(x), \label{FTv-inv}
\end{eqnarray}
where $a_{n}=(-1)^{n}b_{n}$. This result is derived from the fact
that the $2n$-th derivative of the $\delta$-function can be
expressed as
$\delta^{(2n)}(x)=\frac{1}{2\pi}\int_{-\infty}^{\infty}(-1)^{n}k^{2n}e^{-\mathrm{i}kx}$dk.

Let us now consider a Gaussian type potential. The Fourier transform
of the Gaussian function is still a Gaussian function and is given
by
\begin{equation}
F\left[\frac{1}{\sqrt{2\pi\alpha^{2}}}\exp\left(-\frac{x^{2}}{2\alpha^{2}}\right)\right]=\exp\left(-\frac{\alpha^{2}
k^{2}}{2}\right). \label{FT-Gaussian}
\end{equation}
The Taylor expansion of the right-hand side of
Eq.~\eqref{FT-Gaussian} at $k=0$ is
\begin{equation}
\exp\left(-\frac{\alpha^{2}
k^{2}}{2}\right)=\sum_{n=0}^{\infty}(-1)^{n}\frac{1}{n!}\left(\frac{\alpha^{2}}{2}\right)^{n}k^{2n}.
\end{equation}
From Eqs.~\eqref{FTv} and \eqref{FTv-inv}, we deduce that
\begin{equation}
v(x)=\frac{1}{\sqrt{2\pi\alpha^{2}}}\exp\left(-\frac{x^{2}}{2\alpha^{2}}\right)
=\sum_{n=0}^{\infty}\frac{1}{n!}\left(\frac{\alpha^{2}}{2}\right)^{n}\delta^{(2n)}(x).
\end{equation}

It seems a little odd at first glance that an analytic function can
be written in the form of an infinite sum of generalized functions.
We emphasize that this equality does not hold at isolated points,
i.e., we cannot, for instance, say that the equality holds at the
point $x_{0}$. But one can convince oneself that it holds whenever
we consider $v(x)$ as a continuous linear functional that associates
every function $\psi(x)$ which vanishes outside some bounded region
and has continuous derivatives of all orders, a real number
$(v,\psi)$. Mathematically, $v(x)$ is considered a functional in the
sense that $(v,\psi)=\int_{R}v(x)\psi(x)dx$ where the integration is
performed over the real line for this instance. One can also check
the validity of the expansion $v(x)$ in terms of a linear
combination of $\delta^{(2n)}(x)$, denoted as $v_{\delta}(x)$, when
$\psi(\mathbf{x})=\sum_{P}A_{\sigma_{1}\ldots\sigma_{N}}(P|Q)\exp(\mathrm{i}\sum_{j=1}^{N}k_{P_{j}}x_{Q_{j}})$
is the Bethe ansatz wavefunction, by comparing the expressions of
$(v,\psi)$ and $(v_{\delta},\psi)$. In Appendix~\ref{Proof}, we
verify the claim that $(v,\psi)=(v_{\delta},\psi)$.

With this expression for the potential $v(x)$ and after verifying
that $(v,\psi)=(v_{\delta},\psi)$, we can re-write the Hamiltonian
in Eq.~\eqref{Ham1} as
\begin{equation}
\mathcal{H}=-\sum_{j=1}^{N}\frac{\partial^{2}}{\partial
x_{j}^{2}}+2c\sum_{j<l}\sum_{n=0}^{\infty}\frac{1}{n!}
\left(\frac{\alpha^{2}}{2}\right)^{n}\delta^{(2n)}(x_{j}-x_{l})-\frac{H}{2}(N_{\uparrow}-N_{\downarrow}).
\label{Ham2}
\end{equation}
Following Gutkin's work \cite{Gutkin1987}, we can show that this
Hamiltonian is integrable. The boundary condition imposed by
Eq.~\eqref{Ham2} (derived in detail in Appendix~\ref{Scatter}) is
\begin{eqnarray}
\nonumber &&\left(\frac{\partial}{\partial
x_{j+1}}-\frac{\partial}{\partial
x_{j}}\right)\psi|_{x_{j}=x_{j+1}^{+}}-\left(\frac{\partial}{\partial
x_{j+1}}-\frac{\partial}{\partial
x_{j}}\right)\psi|_{x_{j}=x_{j+1}^{-}} \\ &&\quad
=2C\sum_{n=0}^{\infty}
\frac{1}{n!}\left(\frac{\alpha^{2}}{8}\right)^{n}
\left(\frac{\partial}{\partial x_{j+1}}-\frac{\partial}{\partial
x_{j}}\right)^{2n}\psi|_{x_{j}=x_{j+1}}. \label{boundaryCondition}
\end{eqnarray}
Here the interaction strength $C$ is now a $d\times d$ matrix, where
$d$ represents the number of internal energy levels. More
explicitly, $C=cI^{d}$ where $I^{d}$ is a $d\times d$ identity
matrix. The superscripts $+$ and $-$ on the position of the
$(j+1)$-th particle $x_{j}$ have the meaning that $x_{j+1}^{+(-)}$
is infinitesimally greater (or smaller) than $x_{j+1}$. This
boundary condition is a specific case of the ones derived in
Refs.~\cite{Ord1989,Lai1995} for velocity dependent
$\delta$-function potentials. To compute the matching coefficients
$A(\lambda,\mu)$ and $B(\lambda,\mu)$ that are found in
Ref.~\cite{Gutkin1987}, we assume that the wavefunctions before
collision and after collision are
\begin{eqnarray}
\psi|_{x_{j}=x_{j+1}^{-}} &=& e^{\mathrm{i}(\lambda x_{j}+\mu
x_{j+1})},
\\
\psi|_{x_{j}=x_{j+1}^{+}} &=& A(\lambda,\mu)e^{\mathrm{i}(\lambda
x_{j}+\mu x_{j+1})}+B(\lambda,\mu)e^{\mathrm{i}(\mu x_{j}+\lambda
x_{j+1})}.
\end{eqnarray}

Next, we substitute these wavefunctions into
Eq.~\eqref{boundaryCondition} and use Proposition 1 in
Ref.~\cite{Gutkin1987}, i.e., $A(\lambda,\mu)+B(\lambda,\mu)=1$,
which states that there are only two possible plane wave solutions after
collision. These are either where (i) the momenta of scattering particles
are interchanged, or (ii) the momenta of scattering particles
are left unchanged, with the sum of their probabilities 
equal to 1. This yields the solutions for $A(\lambda,\mu)$ and
$B(\lambda,\mu)$, i.e.,
\begin{eqnarray}
A(\lambda,\mu) &=&
\frac{(\lambda-\mu)-\mathrm{i}C\sum_{n=0}^{\infty}\frac{1}{n!}\left(-\frac{\alpha^{2}}{8}\right)^{n}(\lambda-\mu)^{2n}}{\lambda-\mu},
\\ B(\lambda,\mu) &=&
\frac{\mathrm{i}C\sum_{n=0}^{\infty}\frac{1}{n!}\left(-\frac{\alpha^{2}}{8}\right)^{n}(\lambda-\mu)^{2n}}{\lambda-\mu}.
\end{eqnarray}

From Theorem 2(b) in Ref.~\cite{Gutkin1987}, the symmetric Bethe
ansatz, i.e., Bethe's hypothesis for a system of bosons, is
satisfied since we have found a pair of commuting matching
coefficients $A(\lambda,\mu)$ and $B(\lambda,\mu)$ for any matrix
$C=cI^{d}$. Hence we have shown that this model is BA integrable.
The $N$ particle symmetric wavefunction can then be expressed as
\begin{equation}
\psi(x_{Q_{1}}\ll x_{Q_{2}}\ll\ldots\ll x_{Q_{N}})
=\sum_{P}A_{\sigma_{1}\ldots\sigma_{N}}(P|Q)
\exp\left(\mathrm{i}\sum_{j=1}^{N}k_{P_{j}}x_{Q_{j}}\right).
\end{equation}
This wavefunction is a superposition of plane waves with different
amplitudes $A_{\sigma_{1}\ldots\sigma_{N}}(P|Q)$ (not to be confused
with the coefficient $A(\lambda,\mu)$) where $P$ and $Q$ are
permutations of the set of integers $\{1,2,\ldots,N\}$. Each plane
wave is characterized by the permutation $P$ of wavenumbers
$\{k_{j}\}$, therefore the sum contains $N!$ terms. Here
$\sigma_{j}$'s represent the spin coordinates.

It should be noted that the simple procedure of replacing an analytic function by a linear
combination of $2n$-th order derivatives of the $\delta$-function
may lead one to think that any Hamiltonian with a pairwise
interaction potential which is an even function can be exactly
solved via the ABA. However, this is not true. The BA integrability
conditions met by the Gaussian function is actually quite
restrictive. First of all, any non-local potential we choose has to
be well-behaved, smooth and an even function. Secondly, it has to
vanish quickly as a function of the distance between neighboring
particles in order for us to make use of the ABA. Thirdly, the
Gaussian function is unique in the sense that it satisfies both
previous conditions, and can still be reduced to a $\delta$-function
as its width vanishes to zero. This third point enables us to make
sure our results reduce to the Lieb-Liniger case in the limit
$\alpha\rightarrow 0$, which is a necessary condition. These three
points eliminate many candidates for a choice of pairwise
interaction potential. In Appendix \ref{YangYang}, we show that for
the case where $T=0$, there exists a unique solution for the Bethe
roots, and that they are good quantum numbers.

\section{The Ground State}
\label{GroundState} The scattering matrix and the ABA equations for
this model are derived in Appendix \ref{Scatter} and Appendix
\ref{BetheAnsatz}. The ABA equations are given by
\begin{equation}
\exp(\mathrm{i}k_{j}L)=
-\prod_{l=1}^{N}\frac{k_{j}-k_{l}+\mathrm{i}c'(k_{j}-k_{l})}{k_{j}-k_{l}-\mathrm{i}c'(k_{j}-k_{l})}
\prod_{i=1}^{M}\frac{k_{j}-\lambda_{i}-\mathrm{i}c'(k_{j}-\lambda_{i})}{k_{j}-\lambda_{i}},
\qquad j=1,\ldots,N, \label{BA1}
\end{equation}
\begin{equation}
\prod_{l=1}^{N}\frac{\lambda_{i}-k_{l}+\mathrm{i}c'(\lambda_{i}-k_{l})}{\lambda_{i}-k_{l}}
=-\prod_{j=1}^{M}\frac{\lambda_{i}-\lambda_{j}+\mathrm{i}c'(\lambda_{i}-\lambda_{j})}
{\lambda_{i}-\lambda_{j}-\mathrm{i}c'(\lambda_{i}-\lambda_{j})},
\qquad i=1,\ldots,M. \label{BA2}
\end{equation}
where the effective interaction strength
$c'(u)=ce^{-\alpha^{2}u^{2}/8}$ is given in Eq.~\eqref{c'}. Here,
$M$ denotes the number of spin-down bosons in a system where the
vacuum state (initial reference state) consists of $N$ spin-up
bosons. The rapidities for the spin degrees of freedom are given by
$\{\lambda_{i}\}$.

When $T=0$ there are no strings involved in the solution for
$\{\lambda_{i}\}$, i.e., all $\lambda_{i}$'s are real. Taking the
logarithm of the ABA equations gives
\begin{equation}
k_{j}L=2\pi
I_{j}-\sum_{l=1}^{N}\theta\left(\frac{k_{j}-k_{l}}{c'(k_{j}-k_{l})}\right)+\frac{1}{2}\sum_{i=1}^{M}
\theta\left(\frac{k_{j}-\lambda_{i}}{c'(k_{j}-\lambda_{i})}\right)+\sum_{i=1}^{M}
\ln\sqrt{1+\left[\frac{c'(k_{j}-\lambda_{i})}{k_{j}-\lambda_{i}}\right]^{2}},
\label{logBA1}
\end{equation}
\begin{equation}
\frac{1}{2}\sum_{l=1}^{N}\theta\left(\frac{\lambda_{i}-k_{l}}{c'(\lambda_{i}-k_{l})}\right)
-\sum_{l=1}^{N}\ln\sqrt{1+\left[\frac{c'(\lambda_{i}-k_{l})}{\lambda_{i}-k_{l}}\right]^{2}}
=2\pi
J_{i}+\sum_{j=1}^{M}\theta\left(\frac{\lambda_{i}-\lambda_{j}}{c'\lambda_{i}-\lambda_{j})}\right),
\label{logBA2}
\end{equation}
where $\theta(x)=2\tan^{-1}x$. Here, quantum numbers $I_{j}$ are
integers (half-odd integers) when $N-M/2$ is odd (even) and $J_{i}$
are integers (half-odd integers) when $N/2-M$ is odd (even). Let us
then define the functions $h(k)$ and $j(\lambda)$ to represent
``particles'' when $Lh(k)=2\pi I$ and when $Lj(\lambda)=2\pi J$.
This yields
\begin{eqnarray}
\nonumber h(k) &=&
k+\frac{1}{L}\sum_{l=1}^{N}\theta\left(\frac{k-k_{l}}{c'(k-k_{l})}\right)
-\frac{1}{2L}\sum_{i=1}^{M}
\theta\left(\frac{k-\lambda_{i}}{c'(k-\lambda_{i})}\right)
\\ && -\frac{1}{L}\sum_{i=1}^{M}
\ln\sqrt{1+\left[\frac{c'(k-\lambda_{i})}{k-\lambda_{i}}\right]^{2}},
\\ \nonumber
j(\lambda) &=&
\frac{1}{2L}\sum_{l=1}^{N}\theta\left(\frac{\lambda-k_{l}}{c'(\lambda-k_{l})}\right)
-\frac{1}{L}\sum_{j=1}^{M}\theta\left(\frac{\lambda-\lambda_{j}}{c'(\lambda-\lambda_{j})}\right)
\\ &&
-\frac{1}{L}\sum_{l=1}^{N}\ln\sqrt{1+\left[\frac{c'(\lambda-k_{l})}{\lambda-k_{l}}\right]^{2}}.
\end{eqnarray}

In the thermodynamic limit,
\begin{eqnarray}
\nonumber h(k) &=&
k+\int\theta\left(\frac{k-k'}{c'(k-k')}\right)\rho(k')dk'
-\frac{1}{2}\int\theta\left(\frac{k-\lambda}{c'(k-\lambda)}\right)\sigma(\lambda)d\lambda
\\ &&
-\int\ln\sqrt{1+\left[\frac{c'(k-\lambda)}{k-\lambda}\right]^{2}}\sigma(\lambda)d\lambda,
\label{density_h}
\\ \nonumber
j(\lambda) &=&
\frac{1}{2}\int\theta\left(\frac{\lambda-k}{c'(\lambda-k)}\right)\rho(k)dk
-\int\theta\left(\frac{\lambda-\lambda'}{c'(\lambda-\lambda')}\right)\sigma(\lambda')d\lambda'
\\ &&
-\int\ln\sqrt{1+\left[\frac{c'(\lambda-k)}{\lambda-k}\right]^{2}}\rho(k)dk,
\label{density_j}
\end{eqnarray}
where $\rho(k)$ and $\sigma(\lambda)$ are the distribution functions
for charge and spin degrees of freedom, respectively. There are no
``holes'' in the ground state, therefore we can safely take
$\rho^{h}(k)=\sigma^{h}(\lambda)=0$. Define
$\frac{d}{dk}h(k)=2\pi\rho(k)$ and
$\frac{d}{dk}j(\lambda)=2\pi\sigma(\lambda)$. Taking the derivatives
of Eqs.~\eqref{density_h} and \eqref{density_j} finally leads to
expressions for the distribution functions in the form
\begin{eqnarray}
\rho(k) &=& \frac{1}{2\pi}+\int K_{1}(k-k')\rho(k')dk'
-\frac{1}{2}\int K_{1}(k-\lambda)\sigma(\lambda)d\lambda +\int
K_{2}(k-\lambda)\sigma(\lambda)d\lambda, \label{Cdist}
\\
\sigma(\lambda) &=& \frac{1}{2}\int K_{1}(\lambda-k)\rho(k)dk-\int
K_{1}(\lambda-\lambda')\sigma(\lambda')d\lambda'+\int
K_{2}(\lambda-k)\rho(k)dk. \label{Sdist}
\end{eqnarray}
The functions $K_{1}(x)$ and $K_{2}(x)$ are given by
\begin{eqnarray}
K_{1}(x) &=&
\frac{1}{\pi}\frac{c'(x)[1+\frac{\alpha^{2}}{4}x^{2}]}{[c'(x)]^{2}+x^{2}},
\\
K_{2}(x) &=&
\frac{1}{2\pi}\frac{c'(x)}{x}\frac{c'(x)[1+\frac{\alpha^{2}}{4}x^{2}]}{[c'(x)]^{2}+x^{2}}
\equiv\frac{c'(x)}{2x}K_{1}(x).
\end{eqnarray}

\section{The Thermodynamics in the Limits $c\gg 1$ and $\alpha\ll 1$}
\label{Thermodynamics} The model described by the Hamiltonian in
Eq.~\eqref{Ham1} does not include any explicit spin-dependent
forces. Therefore the ground state is ferromagnetic according to a
theorem given by Eisenberg and Lieb \cite{Eisenberg2002}. When the
external magnetic field $H>0$, the ground state is fully populated
by $|\uparrow\rangle$ states which were the reference states that we
used to derive the ABA equations. When $H<0$, all $|\uparrow\rangle$
states will flip into $|\downarrow\rangle$ states. The ferromagnetic
behavior and thermodynamics of the special case $\alpha=0$ has been
studied in literature \cite{Guan2007,Caux2009}.

\begin{figure}
\centering
\begin{tabular}{c}
\epsfig{file=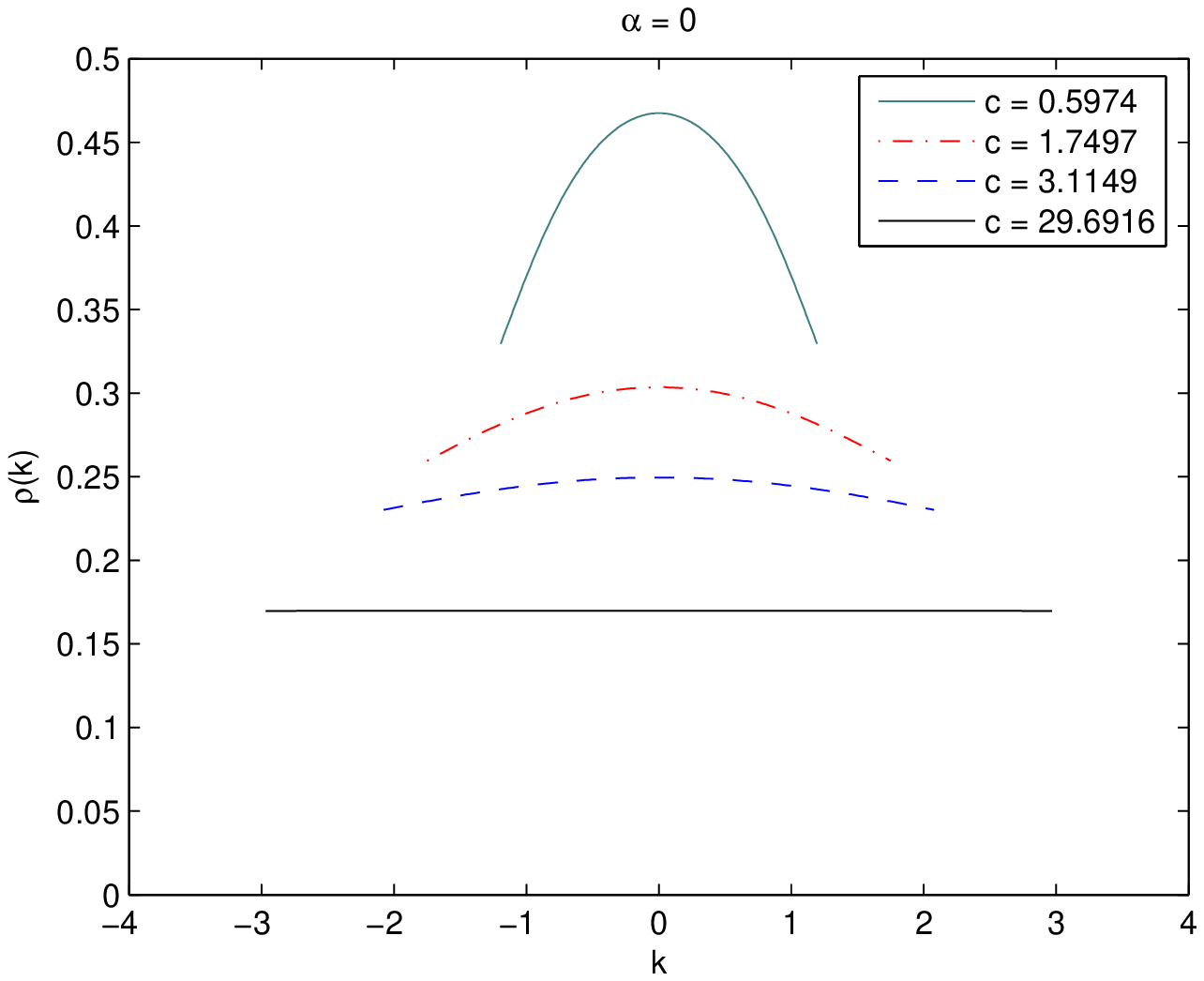,width=0.8\linewidth,clip=} \\
\epsfig{file=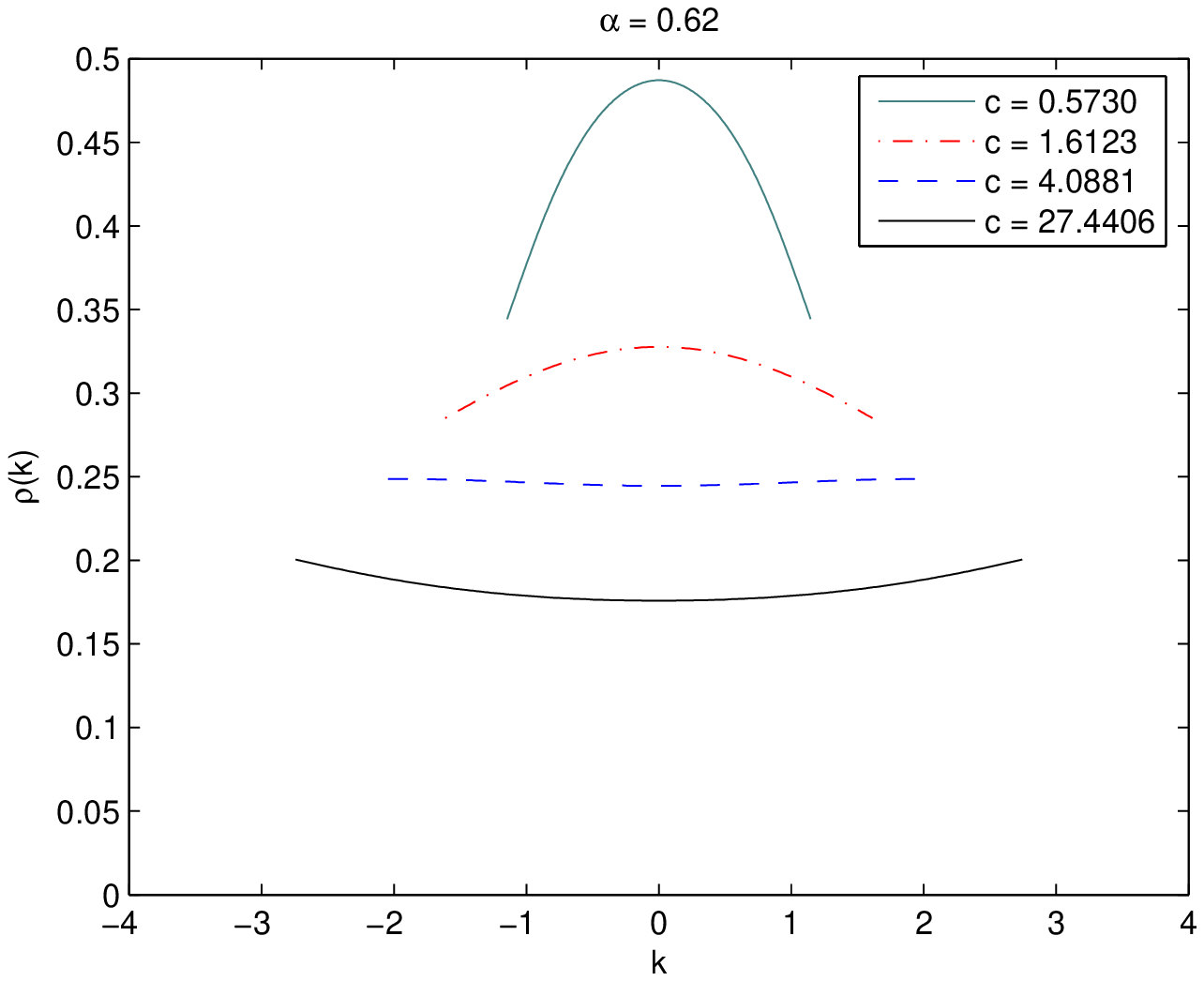,width=0.8\linewidth,clip=}
\end{tabular}\caption{(Color online) 
Plots of $\rho(k)$ versus $k$ for different values of $c$, with
fixed density $n=1$. The top graph has a value of $\alpha=0$ (where
one recovers the Lieb-Liniger Bose gas) and the bottom graph has a
value of $\alpha=0.62$. All curves are obtained by numerically
solving Eq.~\eqref{rho}.}\label{fig:rho1}
\end{figure}

\begin{figure}
\centering
\begin{tabular}{c}
\epsfig{file=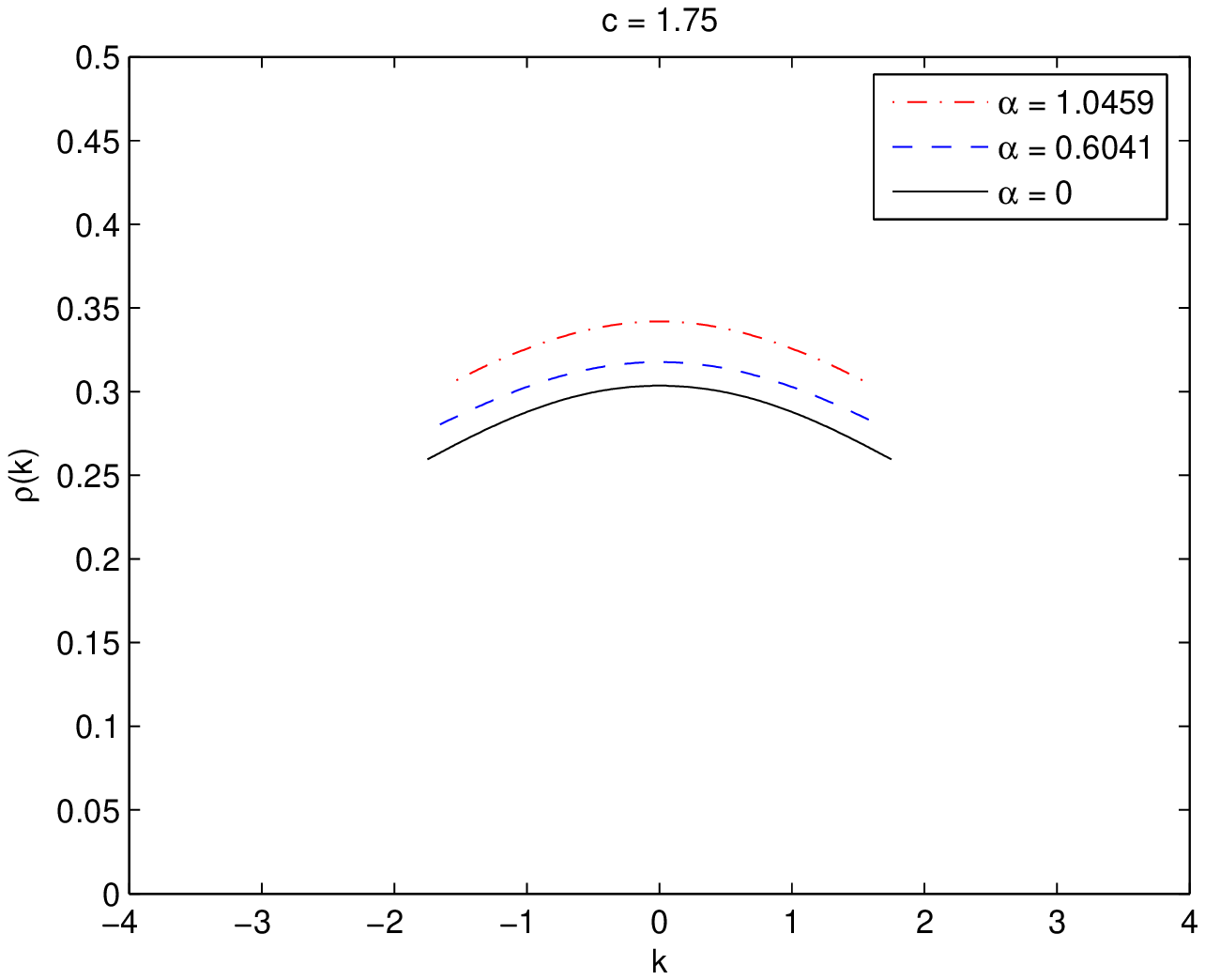,width=0.8\linewidth,clip=} \\
\epsfig{file=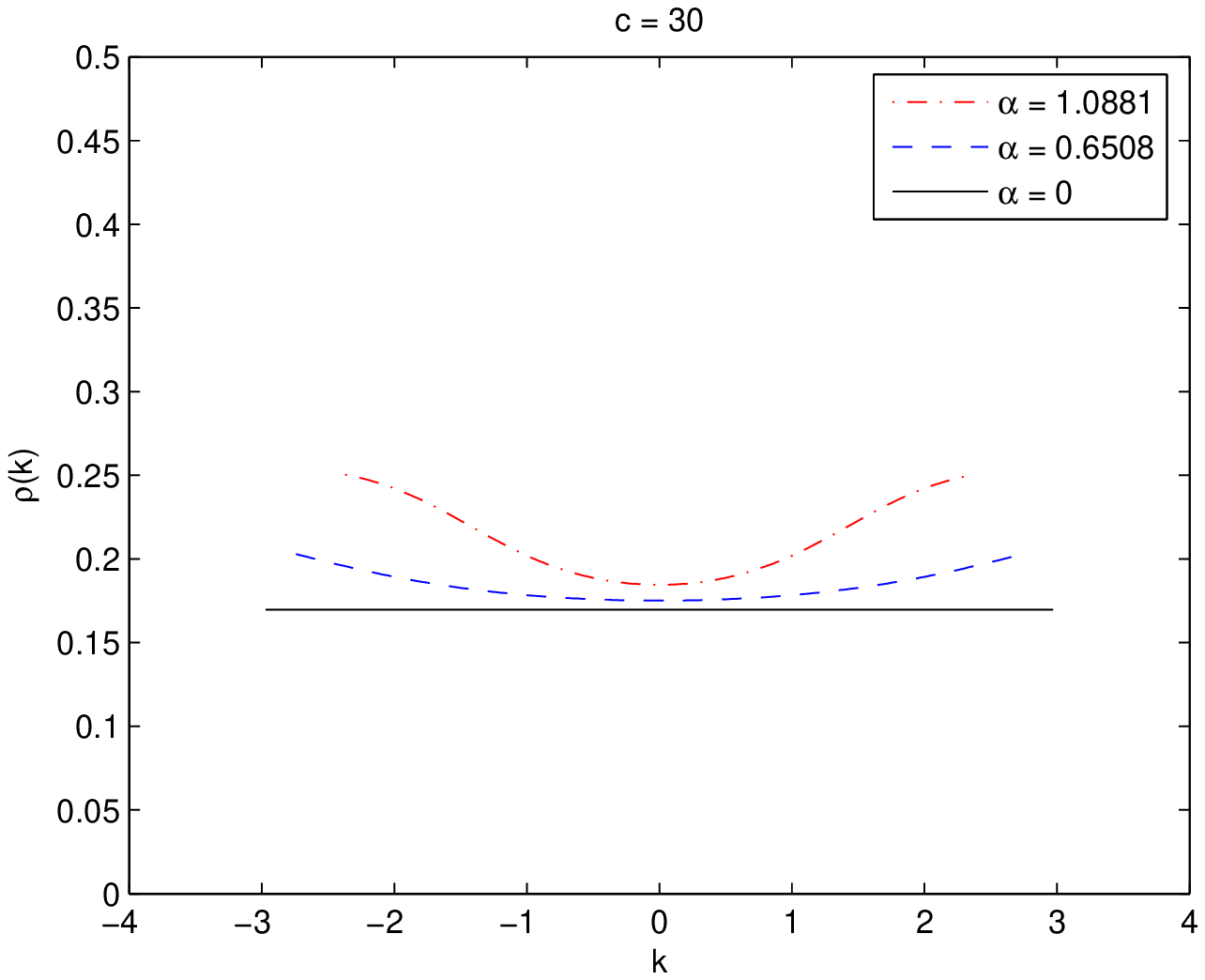,width=0.8\linewidth,clip=}
\end{tabular}\caption{(Color online) 
Numerical plots of $\rho(k)$ versus $k$ for different values of
$\alpha$, with fixed density $n=1$. The top graph has a value of
$c=1.75$ and the bottom graph has a value of $c=30$. All curves are
obtained by numerically solving Eq.~\eqref{rho}.}\label{fig:rho2}
\end{figure}

When $T=0$ and $H>0$, our model reduces to the single component
case. Here $\sigma(\lambda)=0$ since the distribution of
$|\downarrow\rangle$ is zero. Therefore we only have one equation to
solve
\begin{equation}
\rho(k)=\frac{1}{2\pi}+\int_{-Q}^{Q}\frac{1}{\pi}
\frac{c'(k-k')[1+\frac{\alpha^{2}}{4}(k-k')^{2}]}{[c'(k-k')]^{2}+(k-k')^{2}}\rho(k')dk',
\label{rho}
\end{equation}
where $\pm Q$ are the ``Fermi'' points. In FIG.~\ref{fig:rho1} and
FIG.~\ref{fig:rho2}, we plot $\rho(k)$ versus $k$ for different
values of $c$ and $\alpha$ by numerically solving Eq.~\eqref{rho}.
In both figures, we consider values of $\alpha$ and $c$ that are
beyond the ABA regime, i.e., values that are outside the limits
$\alpha\ll 1$ and $c\gg 1$. This is done so that we can more easily
visualize how the distribution function $\rho(k)$ varies as both
parameters vary. We stress that the curves in FIG.~\ref{fig:rho1}
and FIG.~\ref{fig:rho2} become less accurate as $\alpha$ tends to
larger values or as $c$ tends to smaller values. It is clear from
the figures that as the interaction width $\alpha$ increases, the
distribution of quasimomenta $k$ become more centered around the
origin. This is because the increase in overlap between single
particle wavefunctions causes the system to behave more and more
like a Bose-Einstein condensate where the quasimomenta of particles
occupy a smaller region in momentum space.

Using the relations $n=\int_{-Q}^{Q}\rho(k)dk$ and
$E/L=\int_{-Q}^{Q}k^{2}\rho(k)dk$, we can approximate $\rho(k)$ by
using Taylor's expansion to get
\begin{eqnarray}
\nonumber \rho(k) &=& \frac{1}{2\pi}+\int_{-Q}^{Q}\frac{1}{\pi}
\frac{ce^{-\alpha^{2}(k-k')^{2}/8}[1+\frac{\alpha^{2}}{4}(k-k')^{2}]}{c^{2}e^{-\alpha^{2}(k-k')^{2}/4}+(k-k')^{2}}\rho(k')dk'
\\ &=& \nonumber \frac{1}{2\pi}+\frac{1}{\pi
c}\int_{-Q}^{Q}\left(1+\frac{3\alpha^{2}}{8}(k-k')^{2}+\frac{5\alpha^{4}}{128}(k-k')^{4}-\ldots\right)\rho(k')dk'
\\ &=& \nonumber \frac{1}{2\pi}+\frac{1}{\pi c}\left(1+\frac{3\alpha^{2}}{8}k^{2}
+\frac{5\alpha^{4}}{128}k^{4}\right)\int_{-Q}^{Q}\rho(k')dk'+\frac{1}{\pi
c}\left(\frac{3\alpha^{2}}{8}+\frac{15\alpha^{4}}{64}k^{2}\right)\int_{-Q}^{Q}k'^{2}\rho(k')dk'
\\ && \nonumber +\frac{5\alpha^{4}}{128\pi
c}\int_{-Q}^{Q}k'^{4}\rho(k')dk'+\ldots \\ &=& \nonumber
\frac{1}{2\pi}+\frac{n}{\pi c}\left(1+\frac{3\alpha^{2}}{8}k^{2}
+\frac{5\alpha^{4}}{128}k^{4}\right)+\frac{3E\alpha^{2}}{8\pi
Lc}\left(1+\frac{5\alpha^{2}}{8}k^{2}\right)
+\frac{Q^{5}\alpha^{4}}{128\pi^{2}c}\left(1+\frac{2n}{c}\right)
\\ && +O\left(\frac{1}{c^{3}}\right)+O\left(\alpha^{6}\right).
\end{eqnarray}
The expression $\int_{-Q}^{Q}k'^{4}\rho(k')dk'$ was evaluated by
substituting the dominant terms in $\rho(k')$ into the integral,
which gave
\begin{equation}
\int_{-Q}^{Q}k^{4}\rho(k)dk\approx
\int_{-Q}^{Q}k^{4}\left(\frac{1}{2\pi}+\frac{n}{\pi c}\right)dk
=\frac{Q^{5}}{5\pi}\left(1+\frac{2n}{c}\right).
\end{equation}

To find an expression for the Fermi point $Q$, we evaluate the
integral
\begin{eqnarray}
\nonumber n &=& \int_{-Q}^{Q}\rho(k)dk \\ &\approx& \nonumber
\frac{Q}{\pi}\left[1+\frac{2n}{c}+\frac{3E\alpha^{2}}{4Lc}+\frac{nQ^{2}\alpha^{2}}{4c}+\frac{Q^{5}\alpha^{4}}{64\pi
c}\left(1+\frac{2n}{c}\right)+\frac{5EQ^{2}\alpha^{4}}{32Lc}+\frac{nQ^{4}\alpha^{4}}{64c}\right].
\end{eqnarray}
Hence
\begin{eqnarray}
\nonumber Q &=& \pi
n\left[1-\frac{2n}{c}\left(1-\frac{2n}{c}\right)-\frac{\pi^{2}n^{3}\alpha^{2}}{4c}\left(1-\frac{8n}{c}\right)
-\frac{3E\alpha^{2}}{4Lc}\left(1-\frac{4n}{c}-\frac{3E\alpha^{2}}{4Lc}\right)\right.
\\ && \left.-\frac{\pi^{4}n^{5}\alpha^{4}}{32c}\left(1-\frac{14n}{c}\right)
-\frac{5\pi^{2}n^{2}E\alpha^{4}}{32Lc}\left(1-\frac{44n}{5c}\right)\right]
+O\left(\frac{1}{c^{3}}\right)+O\left(\alpha^{6}\right).
\end{eqnarray}

The ground state energy per unit length of the system is given by
\begin{eqnarray}
\nonumber\frac{E}{L} &=& \int_{-Q}^{Q}k^{2}\rho(k)dk \\
&=& \nonumber
\frac{Q^{3}}{3\pi}\left[1+\frac{2n}{c}+\frac{Q^{3}\alpha^{2}}{4\pi
c}\left(1+\frac{2n}{c}\right)
+\frac{9nQ^{2}\alpha^{2}}{20c}+\frac{Q^{6}\alpha^{4}}{16\pi^{2}c^{2}}+\frac{7Q^{5}\alpha^{4}}{64\pi
c}\left(1+\frac{106n}{35c}\right)\right.
\\ &&
\left.+\frac{15nQ^{4}\alpha^{4}}{448c}\right]+O\left(\frac{1}{c^{3}}\right)+O\left(\alpha^{6}\right).
\end{eqnarray}
Substituting $Q$ into $E/L$ and collecting similar terms yields
\begin{eqnarray}
\nonumber \frac{E}{L} &=&
\frac{1}{3}\pi^{2}n^{3}\left[1-\frac{4}{\gamma}\left(1-\frac{3}{\gamma}\right)
-\frac{4\pi^{2}n^{2}\alpha^{2}}{5\gamma}\left(1-\frac{10}{\gamma}\right)
-\frac{3\pi^{4}n^{4}\alpha^{4}}{28\gamma}\left(1+\frac{21}{10\gamma}\right)\right]
\\ && +O\left(\frac{1}{\gamma^{3}}\right)+O\left(\alpha^{6}\right).
\label{gsEnergy}
\end{eqnarray}
where $\gamma=c/n$. With this expression for $E/L$, the ``Fermi''
points can be written explicitly as
\begin{equation}
Q=\pi n\left[1-\frac{2}{\gamma}\left(1-\frac{2}{\gamma}\right)
-\frac{\pi^{2}n^{2}\alpha^{2}}{2\gamma}\left(1-\frac{8}{\gamma}\right)
-\frac{\pi^{4}n^{4}\alpha^{4}}{12\gamma}\left(1-\frac{82}{5\gamma}\right)\right]
+O\left(\frac{1}{\gamma^{3}}\right)+O\left(\alpha^{6}\right).
\end{equation}

\begin{figure}
\centering \epsfig{file=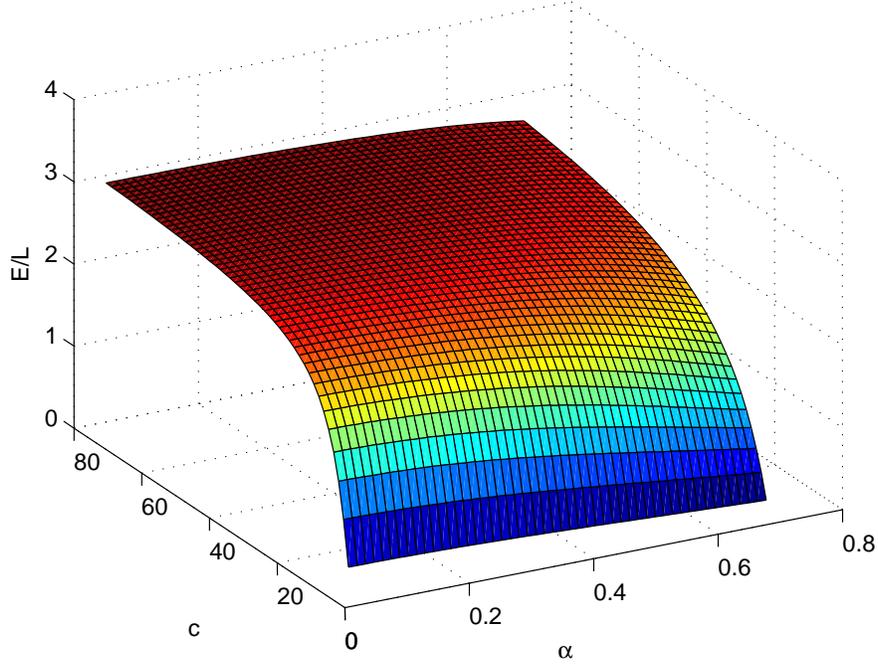,width=0.8\linewidth,clip=}
\caption{(Color online) Plot of the ground state energy per unit
length $E/L$ versus the interaction width $\alpha$ and the
interaction strength $c$ for a fixed density $n=1$. The surface is
generated by numerically solving the equation
$E/L=\int_{-Q}^{Q}k^{2}\rho(k)dk$.}\label{fig:GroundState}
\end{figure}

With the expression for the ground state energy, the chemical
potential can be derived using the relation
\begin{eqnarray}
\nonumber \mu &=& \frac{\partial}{\partial
n}\left(\frac{E}{L}\right)
\\ &=& \nonumber
\pi^{2}n^{2}\left[1-\frac{16}{3\gamma}\left(1-\frac{15}{4\gamma}\right)-\frac{8\pi^{2}n^{2}\alpha^{2}}{5\gamma}
\left(1-\frac{35}{3\gamma}\right)-\frac{2\pi^{4}n^{4}\alpha^{4}}{7\gamma}\left(1+\frac{189}{80\gamma}\right)\right]
\\ &&
+O\left(\frac{1}{\gamma^{3}}\right)+O\left(\alpha^{6}\right).\label{mu}
\end{eqnarray}

\begin{figure}
\centering
\epsfig{file=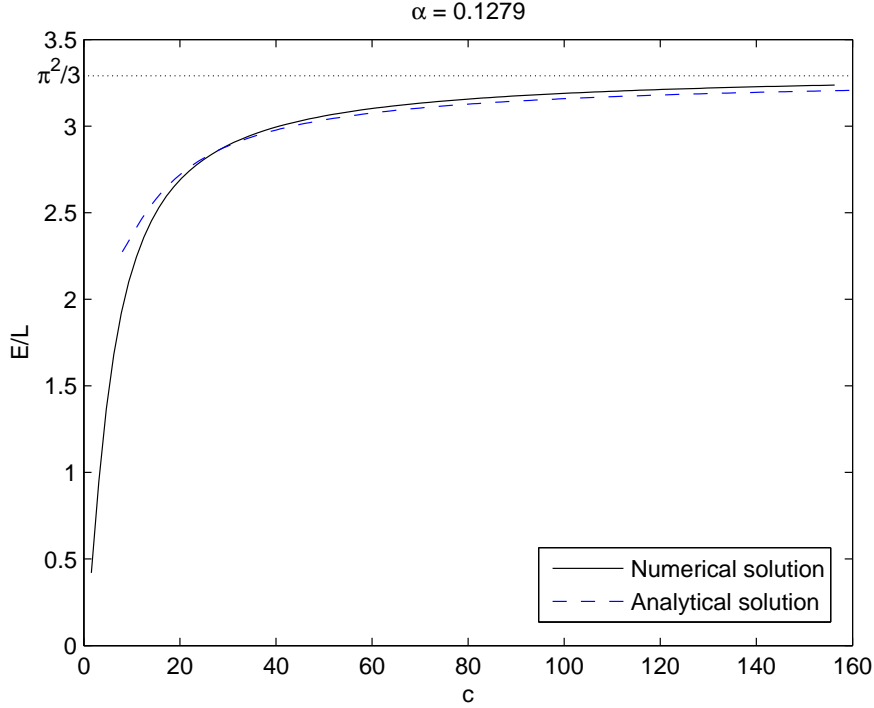,width=0.8\linewidth,clip=}
\caption{(Color online) Comparison between the analytical results
and the numerical results for the ground state energy per unit
length $E/L$ versus $c$ with $\alpha=0.1279$ and
$n=1$.}\label{fig:comparisonC}
\end{figure}

The ground state energy is also calculated numerically for different
values of $\alpha$ and $c$ by using $\rho(k)$ in Eq.~\eqref{rho} and
the definition $E/L=\int_{-Q}^{Q}k^{2}\rho(k)dk$. We thus show a
plot of $E/L$ versus $\alpha$ and $c$ in FIG.~\ref{fig:GroundState}.
As $c$ tends to infinity, the ground state energy will approach
$\pi^{2}n^{3}/3$ as predicted by our analytical results. In
FIG.~\ref{fig:comparisonC}, we compare our analytical solution given
in Eq.~\eqref{gsEnergy} with the numerical solution for the ground
state energy per unit length $E/L$ when $\alpha=0.1279$ and $n=1$.
It is clear they both agree well when $c$ is large.

\section{Local Density Approximation}
\label{Local} In this section, we explore the axial density when the
system is confined by an external harmonic trapping potential. So
far our application of the ABA to solve this model has been limited
to the case where there is no external confinement. When an external
confinement is applied, the model is no longer exactly solvable.
However, if the external trapping potential varies slowly enough,
the local density approximation (LDA) \cite{Dunjko2001} can be
applied to analyze the density profiles in a harmonic trap.

\begin{figure}
\centering \epsfig{file=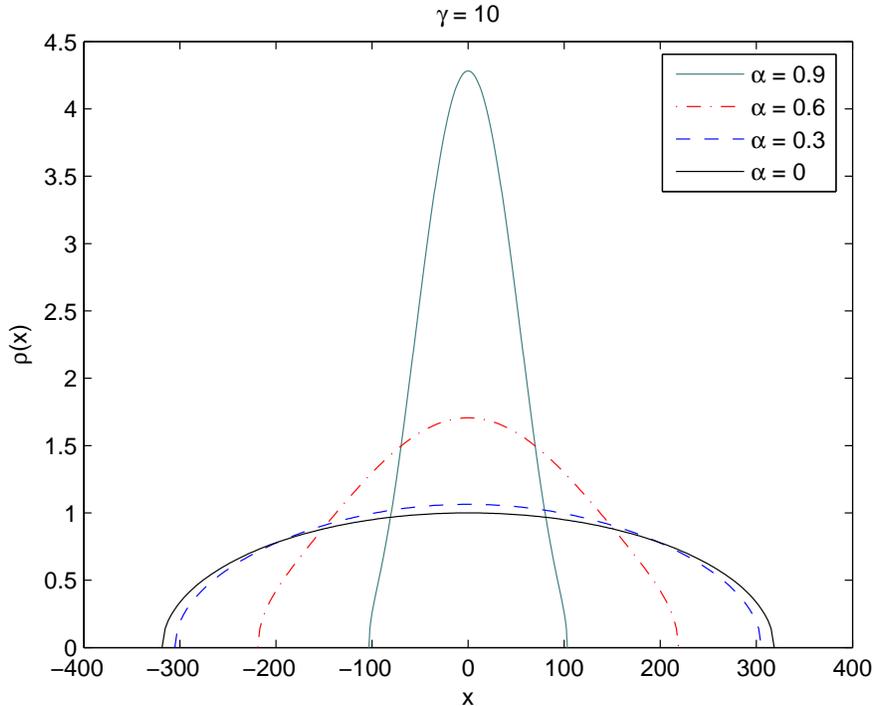,width=0.8\linewidth,clip=}
\caption{(Color online) Axial density profiles from the local
density approximation for different values of $\alpha$. Here
$\gamma=10$, total particle number $N=1000$, and the density at the
center of the trap is taken to be $n(0)=1$.}\label{fig:LDA}
\end{figure}

In the LDA, the chemical potential varies along the axial direction
$x$ according to the equation
\begin{equation}
\mu(x)=\mu(0)-\frac{m\omega^{2}x^{2}}{2}.\label{LDA}
\end{equation}
Using the result in Eq.~\eqref{mu}, we then have
\begin{eqnarray}
\nonumber \mu(0)-\frac{m\omega^{2}x^{2}}{2} &=&
\pi^{2}n(x)^{2}\left[1-\frac{16}{3\gamma}\left(1-\frac{15}{4\gamma}\right)-\frac{8\pi^{2}n(x)^{2}\alpha^{2}}{5\gamma}
\left(1-\frac{35}{3\gamma}\right)\right.
\\ &&
\left.-\frac{2\pi^{4}n(x)^{4}\alpha^{4}}{7\gamma}\left(1+\frac{189}{80\gamma}\right)\right].
\end{eqnarray}
Solving this equation for $n(x)$ gives
\begin{eqnarray}
\nonumber n(x) &=&
n(0)\sqrt{1-\frac{x^{2}}{R^{2}}}\left[1+\frac{4\pi^{2}n(0)^{2}\alpha^{2}}{5\gamma}\left(1-\frac{19}{3\gamma}\right)
\left(1-\frac{x^{2}}{R^{2}}\right)\right.
\\ && \left.+\frac{\pi^{4}n(0)^{4}\alpha^{4}}{7\gamma}\left(1+\frac{28051}{1200\gamma}\right)
\left(1-\frac{x^{2}}{R^{2}}\right)^{2}\right],
\end{eqnarray}
where
\begin{equation}
n(0)=\frac{1}{\pi}\sqrt{\frac{\mu(0)}{1-\frac{16}{3\gamma}(1-\frac{15}{4\gamma})}},
\end{equation}
and
\begin{equation}
R^{2}=\frac{2\mu(0)}{m\omega^2}.
\end{equation}
To obtain the density profiles, we solve the integral
\begin{equation}
\int_{-R}^{R}n(x)dx=N
\end{equation}
numerically with total number $N=1000$ particles and particle
density $n(0)=1$ at the center of the trap.

In FIG.~\ref{fig:LDA}, we show the axial density profiles for
different values of $\alpha$. As the interaction width $\alpha$
increases, the particles become more concentrated at the center of
the trap in a way analogous to a Bose-Einstein condensate.

\section{Conclusion}
\label{Conc} In this paper, we studied a system of interacting
$SU(2)$ spinor bosons in one-dimension with finite range Gaussian
potential. Using Gutkin's argument \cite{Gutkin1987}, this model is
shown to be exactly solvable. We applied the asymptotic Bethe ansatz
to solve this model when the interaction width $\alpha$ is much
smaller than the inter-particle separation $|x_{i}-x_{j}|$. The
Bethe ansatz equations were derived in Eqs.~\eqref{BA1} and
\eqref{BA2} through the quantum inverse scattering method. We went
on to derive the particle distribution functions for the charge and
spin degrees of freedom in Eqs.~\eqref{Cdist} and \eqref{Sdist}. In
the limits $c\gg 1$, $\alpha\ll 1$ and $H>0$, we derived the ground
state energy \eqref{gsEnergy} and chemical potential \eqref{mu} for
the system. The spin independent interaction leads to a
ferromagnetic ground state. Our analytical results were shown to be
consistent with the exact numerical results from the asymptotic
Bethe ansatz equations. Finally, we applied the local density
approximation to analyze the density profiles of the system in an 
harmonic trapping potential. From our results, we showed that an
increase in interaction width $\alpha$ causes the spatial and
momentum density profiles of the system to more closely resemble
that of a Bose-Einstein condensate, in the sense that density
profiles are more concentrated around the origin.

This work has been partially supported by the Australian Research Council.

\appendix
\section{Proof of $(v,\psi)=(v_{\delta},\psi)$}
\label{Proof} Given the Bethe ansatz wavefunction
$\psi(\mathbf{x})=\sum_{P}A_{\sigma_{1}\ldots\sigma_{N}}(P|Q)\exp(\mathrm{i}\sum_{j=1}^{N}k_{P_{j}}x_{Q_{j}})$,
it is straightforward to show that
\begin{eqnarray}
\nonumber (v,\psi) &=&
\int_{-\infty}^{\infty}\frac{1}{\sqrt{2\pi\alpha^{2}}}e^{-x_{Q_{i}}^{2}/2\alpha^{2}}\sum_{P}A_{\sigma_{1}\ldots\sigma_{N}}(P|Q)
\exp\left(\mathrm{i}\sum_{j=1}^{N}k_{P_{j}}x_{Q_{j}}\right)dx_{Q_{i}}
\\ \nonumber &=& \sum_{P}A_{\sigma_{1}\ldots\sigma_{N}}(P|Q)\exp\left(\mathrm{i}\sum_{j\neq
i}^{N}k_{P_{j}}x_{Q_{j}}\right)F\left[\frac{1}{\sqrt{2\pi\alpha^{2}}}\exp\left(-\frac{x_{Q_{i}}^{2}}{2\alpha^{2}}\right)\right]
\\ &=& \sum_{P}A_{\sigma_{1}\ldots\sigma_{N}}(P|Q)\exp\left(\mathrm{i}\sum_{j\neq
i}^{N}k_{P_{j}}x_{Q_{j}}\right)\exp\left(-\frac{\alpha^{2}k^{2}_{P_{i}}}{2}\right),
\end{eqnarray}
and
\begin{eqnarray}
\nonumber (v_{\delta},\psi) &=&
\int_{-\infty}^{\infty}\sum_{n=0}^{\infty}\frac{1}{n!}\left(\frac{\alpha^{2}}{2}\right)^{n}\delta^{(2n)}(x_{Q_{i}})
\sum_{P}A_{\sigma_{1}\ldots\sigma_{N}}(P|Q)\exp\left(\mathrm{i}\sum_{j=1}^{N}k_{P_{j}}x_{Q_{j}}\right)dx_{Q_{i}}
\\ \nonumber &=& \sum_{P}A_{\sigma_{1}\ldots\sigma_{N}}(P|Q)\exp\left(\mathrm{i}\sum_{j\neq
i}^{N}k_{P_{j}}x_{Q_{j}}\right)F\left[\sum_{n=0}^{\infty}\frac{1}{n!}\left(\frac{\alpha^{2}}{2}\right)^{n}\delta^{(2n)}(x_{Q_{i}})\right]
\\ &=& \nonumber \sum_{P}A_{\sigma_{1}\ldots\sigma_{N}}(P|Q)\exp\left(\mathrm{i}\sum_{j\neq
i}^{N}k_{P_{j}}x_{Q_{j}}\right)\sum_{n=0}^{\infty}\frac{1}{n!}\left(\frac{\alpha^{2}}{2}\right)^{n}(-1)^{n}k_{P_{i}}^{2n}
\\ &=& \sum_{P}A_{\sigma_{1}\ldots\sigma_{N}}(P|Q)\exp\left(\mathrm{i}\sum_{j\neq
i}^{N}k_{P_{j}}x_{Q_{j}}\right)\exp\left(-\frac{\alpha^{2}k^{2}_{P_{i}}}{2}\right),
\end{eqnarray}
which verifies the claim that $(v,\psi)=(v_{\delta},\psi)$.

\section{Yang-Yang variational principle}
\label{YangYang} Let us focus on repulsive potentials such that
$v(x)$ is positive definite and $c>0$. When $T=0$,
Eqs.~\eqref{logBA1} and \eqref{logBA2} reduce to
\begin{equation}
k_{j}L=2\pi I_{j}-\sum_{l=1}^{N}2
\tan^{-1}\left(\frac{k_{j}-k_{l}}{c\tilde{v}(k_{j}-k_{l})}\right),
\label{fundeq}
\end{equation}
where $I_j$ is an integer and $\tilde{v}(k)$ is the Fourier
transform of $v(x)$. This is the fundamental equation for the Bethe
roots which can be posed as a variational principle as shown by Yang and Yang 
for spinless bosons \cite{YangYang1969}. In order to show that Eq.~\eqref{fundeq} can be
uniquely parameterized, we introduce the action
\begin{equation}
B(k_1,\dots,k_N)=\frac{L}{2}\sum_{j=1}^{N}k_{j}^2-2\pi I_{j}
k_{j}+\sum_{j<l}\Phi(k_j-k_l)
\end{equation}
with
\begin{equation}
\Phi(x)=\int_{0}^{x}2\tan^{-1}\left(\frac{x'}{c\tilde{v}(x')}\right)dx'.
\end{equation}
Then we need to show that Eq. \eqref{fundeq} is given by the minima
condition
\begin{equation}
\frac{\partial B(k_1,\dots,k_N)}{\partial k_j}=0.
\end{equation}

To prove this, we further introduce the $N\times N$ matrix
\begin{eqnarray}
\nonumber B_{jl}=\frac{\partial^2 B}{\partial k_{j}\partial
k_{l}}&=&\delta_{jl}\left[L+2c\sum_{m}\frac{\vartheta(k_{j}-k_{m})}
{c^2\tilde{v}^{2}(k_{j}-k_{m})+(k_{j}-k_{m})^{2}}\right] \\
&&-2c\frac{\vartheta(k_{j}-k_{l})}{c^2\tilde{v}^2(k_{j}-k_{l})+(k_{j}-k_{l})^{2}}
\end{eqnarray}
which is always positive provided that
\begin{equation}
\vartheta(k)=\tilde{v}(k)-k\tilde{v}'(k)>0.\label{condition}
\end{equation}
If that is the case
\begin{equation}
\sum_{lj}u_{l}B_{lj}u_{j}=L\sum_{l}
u_{l}^{2}+\sum_{l<j}c\frac{\vartheta(k_{j}-k_{l})}
{c^2\tilde{v}^2(k_{j}-k_{l})+(k_{j}-k_{l})^2}(u_{j}-u_{l})^2\geq 0,
\end{equation}
for arbitrary real $\{u_{j}\}$. Hence, the solutions of the
fundamental equation exist and can be uniquely parameterized by a
set of integer or half-integer numbers $I_{j}$, as long as
$\vartheta(k)=\tilde{v}(k)-k\tilde{v}'(k)\geq 0$.

We shall exclusively consider such type of potentials. Then, the
Bethe roots are real numbers from Theorem I on p. 11 of Ref. 
\cite{Korepin}. Finally if
$I_l>I_m$ then $k_l>k_m$ and if $I_l=I_m$ then $k_l=k_m$ as long as
$\tan^{-1}\left(k/c\tilde{v}(k)\right)$ increases monotonically with
$k$. For the Gaussian potential,
$\tilde{v}(k)=\exp(-\alpha^{2}k^{2}/2)$, which gives
$\vartheta(k)=\tilde{v}(k)-k\tilde{v}'(k)=\tilde{v}(k)(1+\alpha^{2}k^{2})>0$
for all real $k$. Therefore, there is a unique solution for the BA
equations when the Gaussian potential is used.

\section{Derivation of the Scattering Matrix}
\label{Scatter} We employ the coordinate BA to obtain the scattering
matrix between two particles. This technique is well known, as used
by Yang \cite{Yang1967} in solving the spin-1/2 fermion model. First
consider the region
\begin{equation}
R:\quad 0\ll x_{Q_{1}}\ll\ldots\ll x_{Q_{j}}\ll
x_{Q_{j+1}}\ll\ldots\ll x_{Q_{N}}\ll L.
\end{equation}
Define a wavefunction in $R$ as
\begin{equation}
\psi(\mathbf{x})=\sum_{P}A_{\sigma_{1}\ldots\sigma_{N}}(P|Q)
\exp\mathrm{i}(k_{P_{1}}x_{Q_{1}}+\ldots+k_{P_{j}}x_{Q_{j}}+k_{P_{j+1}}x_{Q_{j+1}}+\ldots+k_{P_{N}}x_{Q_{N}}),
\label{Wfunction1}
\end{equation}
where $\sigma_{j}$s represent the spin coordinates. This
wavefunction is a superposition of plane waves with different
amplitudes $A_{\sigma_{1}\ldots\sigma_{N}}(P|Q)$ where $P$ and $Q$
are permutations of the set of integers $\{1,2,\ldots,N\}$. Each
plane wave is characterized by the permutation $P$ of wavenumbers
$\{k_{j}\}$, therefore the sum contains $N!$ terms.

Consider a new region $R'$ where particles at position $x_{Q_{j}}$
and $x_{Q_{j+1}}$ are interchanged, i.e.,
\begin{equation}
R':\quad 0\ll x_{Q_{1}}\ll\ldots\ll x_{Q_{j+1}}\ll
x_{Q_{j}}\ll\ldots\ll x_{Q_{N}}\ll L.
\end{equation}
In this region, the wavefunction is defined as
\begin{equation}
\psi'(\mathbf{x})=\sum_{P}A_{\sigma_{1}\ldots\sigma_{N}}(P|Q')
\exp\mathrm{i}(k_{P_{1}}x_{Q_{1}}+\ldots+k_{P_{j}}x_{Q_{j+1}}+k_{P_{j+1}}x_{Q_{j}}+\ldots+k_{P_{N}}x_{Q_{N}}).
\label{Wfunction2}
\end{equation}
From the condition that the wavefunction has to be continuous when
$x_{Q_{j}}\rightarrow x_{Q_{j+1}}$, we have the relation
\begin{equation}
A_{\sigma_{1}\ldots\sigma_{N}}(P|Q)+A_{\sigma_{1}\ldots\sigma_{N}}(P'|Q)
=A_{\sigma_{1}\ldots\sigma_{N}}(P|Q')+A_{\sigma_{1}\ldots\sigma_{N}}(P'|Q').
\label{continuity}
\end{equation}
where $P'$ and $Q'$ represent the permutations
$P'=(j\phantom{ab}j+1)P$ and $Q'=(j\phantom{ab}j+1)Q$, i.e., only
the positions of the $j$-th and $(j+1)$-th terms are transposed to
get $P'$ from $P$, and $Q'$ from $Q$.

The $\delta$-function potential gives rise to a jump in the first
derivative of the wavefunction at position $x_{Q_{j}}=x_{Q_{j+1}}$.
This jump can be evaluated by considering the Hamiltonian in the
center of mass frame. In this frame, the new coordinates $X$ and $Y$
are related to the original coordinates $x_{j}$ and $x_{j+1}$ by the
transformation relations
\begin{equation}
X=\frac{x_{j}+x_{j+1}}{2}, \qquad Y=x_{j+1}-x_{j},
\end{equation}
and
\begin{equation}
x_{j}=X-\frac{Y}{2}, \qquad x_{j+1}=X+\frac{Y}{2}.
\end{equation}
Their derivatives are related by
\begin{equation}
\frac{\partial}{\partial x_{j}}=\frac{1}{2}\frac{\partial}{\partial
X}-\frac{\partial}{\partial Y},\qquad \frac{\partial}{\partial
x_{j+1}}=\frac{1}{2}\frac{\partial}{\partial
X}+\frac{\partial}{\partial Y},
\end{equation}
and
\begin{equation}
\frac{\partial}{\partial X}=\frac{\partial}{\partial
x_{j}}+\frac{\partial}{\partial x_{j+1}},\qquad
\frac{\partial}{\partial
Y}=\frac{1}{2}\left(\frac{\partial}{\partial
x_{j+1}}-\frac{\partial}{\partial x_{j}}\right).
\end{equation}
Higher order derivatives can be similarly expressed in a
straightforward manner.

The time-independent Schr\"{o}dinger equation
$\mathcal{H}\psi=E\psi$ in these new coordinates is then given by
\begin{equation}
\left\{-\frac{1}{2}\frac{\partial^{2}}{\partial
X^{2}}-2\frac{\partial^{2}}{\partial
Y^{2}}+2c\sum_{n=0}^{\infty}\frac{1}{n!}\left(\frac{\alpha^{2}}{2}\right)^{n}
\delta^{(2n)}(Y)+\ldots\right\}\psi(X,Y,\mathbf{x'})=E\psi(X,Y,\mathbf{x'}).
\end{equation}
where the new set of coordinates $X$, $Y$ and $\mathbf{x'}$ replace
the old one $\mathbf{x}$. Also, the dimension of $\mathbf{x'}$ is
less than the dimension of $\mathbf{x}$ by two, since we replaced
those two coordinates by $X$ and $Y$. Integrating this equation with
respect to the $Y$ coordinate from $-\epsilon$ to $\epsilon$ and
then taking $\epsilon\rightarrow 0$ gives
\begin{equation}
\left.\frac{\partial\psi}{\partial
Y}\right|_{Y=0^{+}}-\left.\frac{\partial\psi}{\partial
Y}\right|_{Y=0^{-}}=c\sum_{n=0}^{\infty}\frac{1}{n!}\left(\frac{\alpha^{2}}{2}\right)^{n}
\left.\frac{\partial^{2n}\psi}{\partial Y^{2n}}\right|_{Y=0},
\label{deltaCondition}
\end{equation}
where we have repeatedly used integration by-parts to obtain the
right hand side of the equation.

In the new coordinates, the wavefunctions given in
Eqs.~\eqref{Wfunction1} and \eqref{Wfunction2} are explicitly
written as
\begin{equation}
\psi(X,Y,\mathbf{x'})=\sum_{P}A_{\sigma_{1}\ldots\sigma_{N}}(P|Q)
\exp\mathrm{i}\left(\ldots+(k_{P_{j}}+k_{P_{j+1}})X+\frac{1}{2}(k_{P_{j+1}}-k_{P_{j}})Y+\ldots\right),
\label{Wfunction1new}
\end{equation}
and
\begin{equation}
\psi'(X,Y,\mathbf{x'})=\sum_{P}A_{\sigma_{1}\ldots\sigma_{N}}(P|Q')
\exp\mathrm{i}\left(\ldots+(k_{P_{j}}+k_{P_{j+1}})X-\frac{1}{2}(k_{P_{j+1}}-k_{P_{j}})Y+\ldots\right).
\label{Wfunction2new}
\end{equation}

Substituting the wavefunctions defined in Eqs.~\eqref{Wfunction1new}
and \eqref{Wfunction2new} into Eq.~\eqref{deltaCondition}
separately, and then adding both equations together yields the
relation
\begin{eqnarray}
\nonumber
\lefteqn{\frac{\mathrm{i}}{2}(k_{P_{j+1}}-k_{P_{j}})\left[A_{\sigma_{1}\ldots\sigma_{N}}(P|Q)
-A_{\sigma_{1}\ldots\sigma_{N}}(P'|Q)\right]} \\ && \nonumber
+\lefteqn{\frac{\mathrm{i}}{2}(k_{P_{j+1}}-k_{P_{j}})\left[A_{\sigma_{1}\ldots\sigma_{N}}(P|Q')
-A_{\sigma_{1}\ldots\sigma_{N}}(P'|Q')\right]}
\\ && =c\sum_{n=0}^{\infty}\frac{1}{n!}\left(\frac{\alpha^{2}}{2}\right)^{n}\left[\frac{\mathrm{i}}{2}(k_{P_{j+1}}-k_{P_{j}})\right]^{2n}
\left[A_{\sigma_{1}\ldots\sigma_{N}}(P|Q)-A_{\sigma_{1}\ldots\sigma_{N}}(P'|Q)\right].
\label{delta}
\end{eqnarray}
We introduce the transposition operator $T_{i,j}$ which transposes
the $i$th and $j$th spatial coordinates of the wavefunction, i.e.,
\begin{eqnarray}
\nonumber \lefteqn{A_{\sigma_{1}\ldots\sigma_{N}}(\ldots P_{i}
\ldots P_{j}\ldots|\ldots Q_{j} \ldots Q_{i}\ldots)} \\ &&
=\left[T_{i,j}\right]^{\sigma'_{1}\ldots\sigma'_{N}}_{\sigma_{1}\ldots\sigma_{N}}A_{\sigma'_{1}\ldots\sigma'_{N}}(\ldots
P_{i} \ldots P_{j}\ldots|\ldots Q_{i} \ldots Q_{j}\ldots).
\end{eqnarray}
In matrix form, this operator $T_{i,j}$ can be written as
$\left[T_{i,j}\right]^{\sigma'_{1}\ldots\sigma'_{N}}_{\sigma_{1}\ldots\sigma_{N}}
=\pm\delta_{\sigma_{i},\sigma'_{j}}\delta_{\sigma_{j},\sigma'_{i}}\prod_{r\neq
i,j}\delta_{\sigma_{r},\sigma'_{r}}$, i.e.,
$T_{i,j}=\mathcal{P}_{i,j}$ for bosons and
$T_{i,j}=-\mathcal{P}_{i,j}$ for fermions where $\mathcal{P}_{i,j}$
is the permutation operator.

Combining this relation together with Eq.~\eqref{continuity}
transforms Eq.~\eqref{delta} to
\begin{eqnarray}
\nonumber &&
\mathrm{i}(k_{P_{j+1}}-k_{P_{j}})\left[A_{\sigma_{1}\ldots\sigma_{N}}(P|Q)-\left[T_{j,j+1}\right]
^{\sigma'_{1}\ldots\sigma'_{N}}_{\sigma_{1}\ldots\sigma_{N}}A_{\sigma'_{1}\ldots\sigma'_{N}}(P'|Q)\right]
\\ && =c\sum_{n=0}^{\infty}\frac{1}{n!}\left(\frac{\alpha^{2}}{2}\right)^{n}\left[\frac{\mathrm{i}}{2}(k_{P_{j+1}}-k_{P_{j}})\right]^{2n}
\left[A_{\sigma_{1}\ldots\sigma_{N}}(P|Q)+I^{\sigma'_{1}\ldots\sigma'_{N}}_{\sigma_{1}\ldots\sigma_{N}}
A_{\sigma'_{1}\ldots\sigma'_{N}}(P'|Q)\right].
\end{eqnarray}
Rearranging the terms finally gives us an expression which relates
the amplitudes of the wavefunction before and after collision, i.e.,
\begin{eqnarray}
\lefteqn{A_{\sigma_{1}\ldots\sigma_{N}}(P|Q)}\label{transposition1}
\\ \nonumber &&=\left[\frac{\mathrm{i}(k_{P_{j+1}}-k_{P_{j}})T_{j,j+1}
+c\sum_{n=0}^{\infty}\frac{1}{n!}\left(\frac{\alpha^{2}}{2}\right)^{n}\left[\frac{\mathrm{i}}{2}(k_{P_{j+1}}-k_{P_{j}})\right]^{2n}I}
{\mathrm{i}(k_{P_{j+1}}-k_{P_{j}})-c\sum_{n=0}^{\infty}\frac{1}{n!}\left(\frac{\alpha^{2}}{2}\right)^{n}
\left[\frac{\mathrm{i}}{2}(k_{P_{j+1}}-k_{P_{j}})\right]^{2n}}
\right]^{\sigma'_{1}\ldots\sigma'_{N}}_{\sigma_{1}\ldots\sigma_{N}}A_{\sigma'_{1}\ldots\sigma'_{N}}(P'|Q).
\end{eqnarray}
Here $I$ is the identity operator which is included into the
relation so that it can be expressed in matrix form. The general
expression of the scattering matrix is given by the term inside the
square bracket as
\begin{equation}
Y_{i,j}(u)=\frac{\mathrm{i}uT_{i,j} +ce^{-\alpha^{2}u^{2}/8}I}
{\mathrm{i}u-ce^{-\alpha^{2}u^{2}/8}}, \label{scatteringmatrixY}
\end{equation}
which relates any two amplitudes before and after collision between
particles at the $i$th and $j$th position whereby the change in
momentum is $u$. The sums in Eq.~\eqref{transposition1} are the
Taylor expansions of the exponential function given in
Eq.~\eqref{scatteringmatrixY}.

For this model to be integrable, the scattering matrix $Y_{i,j}(u)$
has to obey the Yang-Baxter relations. To see whether this is true,
we shall consider the transposition of two amplitudes through
different paths. Without any loss of generality, consider going from
$A_{123}(123|Q)$ to $A_{321}(321|Q)$ along the two different paths
\begin{eqnarray}
\nonumber A_{123}(123|Q) &=&
[Y_{1,2}(k_{2}-k_{1})]^{213}A_{213}(213|Q) \\
&=& \nonumber
[Y_{1,2}(k_{2}-k_{1})]^{213}[Y_{2,3}(k_{3}-k_{1})]^{231}A_{231}(231|Q)
\\ &=&
[Y_{1,2}(k_{2}-k_{1})]^{213}[Y_{2,3}(k_{3}-k_{1})]^{231}[Y_{1,2}(k_{3}-k_{2})]^{321}A_{321}(321|Q),
\end{eqnarray}
and
\begin{eqnarray}
\nonumber A_{123}(123|Q) &=&
[Y_{2,3}(k_{3}-k_{2})]^{132}A_{132}(132|Q) \\
&=& \nonumber
[Y_{2,3}(k_{3}-k_{2})]^{132}[Y_{1,2}(k_{3}-k_{1})]^{312}A_{312}(312|Q)
\\ &=&
[Y_{2,3}(k_{3}-k_{2})]^{132}[Y_{1,2}(k_{3}-k_{1})]^{312}[Y_{2,3}(k_{2}-k_{1})]^{321}A_{321}(321|Q).
\end{eqnarray}
Since the outcome of both paths is the same, they must be equal to
each other. In general, the scattering matrices satisfy the
Yang-Baxter relations
\begin{eqnarray}
\nonumber Y_{a,b}(u)Y_{c,d}(v) &=& Y_{c,d}(v)Y_{a,b}(u), \\
\nonumber Y_{a,b}(u)Y_{b,c}(u+v)Y_{a,b}(v) &=& Y_{b,c}(v)Y_{a,b}(u+v)Y_{b,c}(u), \\
Y_{a,b}(u)Y_{b,a}(-u) &=& 1. \label{YangBaxterY}
\end{eqnarray}

\section{Derivation of the Bethe Ansatz Equations}
\label{BetheAnsatz}
\subsection{The Quantum Inverse Scattering Method}
We will use the quantum inverse scattering method (QISM)
\cite{Korepin} to derive the ABA equations for this model. On
introducing the operator $R_{i,j}(u)=P_{i,j}Y_{i,j}(u)$ where
$P_{i,j}$ is the permutation matrix, we have the Yang-Baxter
equations in terms of $R_{i,j}(u)$, i.e.,
\begin{equation}
R_{a,b}(u)R_{a,c}(u+v)R_{b,c}(v)=R_{b,c}(v)R_{a,c}(u+v)R_{a,b}(u).
\end{equation}
Notice the difference in subscripts between the above equation and
the second equation in Eq.~\eqref{YangBaxterY}. The $R$-matrices act
on the state space of this $N$ particle system
$V_{N}=\prod_{n=1}^{N}\otimes V_{n}$, i.e., $R_{a,b}(u)$ acts
non-identically on the tensor subspaces $V_{a}$ and $V_{b}$ and
identically on the rest of the subspaces.

Using the Lax representation, we introduce the $L$-operator which
acts on the auxiliary space and a quantum state space, i.e.,
$L_{m}(u)\equiv R_{a,m}(u)$ where $a$ is the auxiliary space and $m$
is the quantum state space. In addition, we also introduce the
interwining operator $\check{R}(u)=\mathcal{P}R(u)$ where the
permutation operator $\mathcal{P}$ has the tensor property on
operators $\mathcal{P}(A\otimes B)\mathcal{P}=B\otimes A$. Hence in
Lax representation, the Yang-Baxter relation becomes
\begin{equation}
\check{R}(u-v)L_{n}(u)\otimes L_{n}(v)=L_{n}(v)\otimes
L_{n}(u)\check{R}(u-v).
\end{equation}

The next step is to introduce the monodromy matrix
$T(u)=L_{N}(u)L_{N-1}(u)\ldots L_{1}(u)$ which is the transition
matrix through the entire ``lattice''. In this form, the Yang-Baxter
relation can be re-written as
\begin{equation}
\check{R}(u-v)T(u)\otimes T(v)=T(v)\otimes T(u)\check{R}(u-v).
\label{YangBaxterT}
\end{equation}

Lastly we introduce the transfer matrix
$\tau(u)=\mathrm{tr}_{a}T(u)$ where the notation $\mathrm{tr}_{a}$
implies that the trace is taken in the auxiliary space. As a
consequence of Eq.~\eqref{YangBaxterT}, there exists a family of
commuting transfer matrices $\tau(u)$, i.e., $[\tau(u),\tau(v)]=0$.
Following the introduction of the operators given above, we can
proceed with our derivation of the ABA equations. As stated earlier,
we are interested in the case where this model has periodic boundary
conditions, i.e.,
\begin{equation}
\psi(x_{1},\ldots,x_{j}=0,\ldots,x_{N})=\psi(x_{1},\ldots,x_{j}=L,\ldots,x_{N}).
\end{equation}
For this condition to hold, the wavefunction defined in
Eq.~\eqref{Wfunction1} has to satisfy
\begin{equation}
A(P_{j},P_{1},\ldots,P_{N}|Q_{j},Q_{1},\ldots,Q_{N})
=\exp(\mathrm{i}k_{j}L)A(P_{1},\ldots,P_{N},P_{j}|Q_{1},\ldots,Q_{N},Q_{j}).
\end{equation}

As a result, we obtain
\begin{eqnarray}
\nonumber\lefteqn{\exp(\mathrm{i}k_{j}L)A_{E}(P|Q)} \\ &&
=R_{j+1,i}(k_{j+1}-k_{j})\ldots
R_{N,j}(k_{N}-k_{j})R_{1,j}(k_{1}-k_{j})\ldots
R_{j-1,j}(k_{j-1}-k_{j})A_{E}(P|Q),
\end{eqnarray}
where $A_{E}(P|Q)$ is the initial amplitude before any
transposition. We can abbreviate this equation as
\begin{equation}
\mathcal{R}_{j}(k_{j})A_{E}(P|Q)=\exp(\mathrm{i}k_{j}L)A_{E}(P|Q),
\label{eigenEquation}
\end{equation}
with the definition
\begin{equation}
\mathcal{R}_{j}(k_{j})=R_{j+1,i}(k_{j+1}-k_{j})\ldots
R_{N,j}(k_{N}-k_{j})R_{1,j}(k_{1}-k_{j})\ldots
R_{j-1,j}(k_{j-1}-k_{j}).
\end{equation}

If we define the monodromy matrix to be
\begin{equation}
T_{N}(u)=L_{N}(k_{N}-u)\ldots L_{2}(k_{2}-u)L_{1}(k_{1}-u),
\label{monodromy}
\end{equation}
the transfer matrix will have the property
\begin{equation}
\tau(u)|_{u=k_{j}}=\mathcal{R}_{j}(k_{j}).
\end{equation}
Hence the eigenvalues of Eq.~\eqref{eigenEquation} coincide with the
eigenvalues of the equation
\begin{equation}
\tau(u)A_{E}(P|Q)=\exp(\mathrm{i}k_{j}L)A_{E}(P|Q)
\label{eigenEquation2}
\end{equation}
at the points $u=k_{j}$ for all $1\leq j\leq N$.

\subsection{The Algebraic Bethe Ansatz}
The $R$-matrix for $SU(2)$ is a $4\times 4$ matrix given by
\begin{equation}
R_{i,j}(u)=\frac{uI-\mathrm{i}c'(u)\mathcal{P}_{i,j}}{u+\textrm{i}c'(u)}=\left(%
\begin{array}{cccc}
  \frac{u-\mathrm{i}c'(u)}{u+\mathrm{i}c'(u)} & 0 & 0 & 0 \\
  0 & \frac{u}{u+\mathrm{i}c'(u)} & -\frac{\mathrm{i}c'(u)}{u+\mathrm{i}c'(u)} & 0 \\
  0 & -\frac{\mathrm{i}c'(u)}{u+\mathrm{i}c'(u)} & \frac{u}{u+\mathrm{i}c'(u)} & 0 \\
  0 & 0 & 0 & \frac{u-\mathrm{i}c'(u)}{u+\mathrm{i}c'(u)} \\
\end{array}%
\right)\equiv\left(
               \begin{array}{cc}
                 a(u) & b(u) \\
                 c(u) & d(u) \\
               \end{array}
             \right),
\end{equation}
where
\begin{equation}
c'(u)=ce^{-\alpha^{2}u^{2}/8}, \label{c'}
\end{equation}
and the matrix representation of the permutation operator is given
by
\begin{equation}
\mathcal{P}_{i,j}=\left(%
\begin{array}{cccc}
  1 & 0 & 0 & 0 \\
  0 & 0 & 1 & 0 \\
  0 & 1 & 0 & 0 \\
  0 & 0 & 0 & 1 \\
\end{array}%
\right).
\end{equation}
Similarly,
\begin{equation}
\check{R}_{i,j}(u)=\frac{u\mathcal{P}_{i,j}-\mathrm{i}c'(u)}{u+\textrm{i}c'(u)}=\left(%
\begin{array}{cccc}
  \frac{u-\mathrm{i}c'(u)}{u+\mathrm{i}c'(u)} & 0 & 0 & 0 \\
  0 & -\frac{\mathrm{i}c'(u)}{u+\mathrm{i}c'(u)} & \frac{u}{u+\mathrm{i}c'(u)} & 0 \\
  0 & \frac{u}{u+\mathrm{i}c'(u)} & -\frac{\mathrm{i}c'(u)}{u+\mathrm{i}c'(u)} & 0 \\
  0 & 0 & 0 & \frac{u-\mathrm{i}c'(u)}{u+\mathrm{i}c'(u)} \\
\end{array}%
\right).
\end{equation}

By choosing the basis for spin-up and spin-down states as
\begin{equation}
|\uparrow\rangle=\left(%
\begin{array}{c}
  1 \\
  0 \\
\end{array}%
\right), \qquad |\downarrow\rangle=\left(%
\begin{array}{c}
  0 \\
  1 \\
\end{array}%
\right),
\end{equation}
we can then act each $2\times 2$ block of the $R$-matrix on the
spin-up basis vector to get
\begin{eqnarray}
a(u)\left(%
\begin{array}{c}
  1 \\
  0 \\
\end{array}%
\right)&=&\frac{u-\mathrm{i}c'(u)}{u+\mathrm{i}c'(u)}\left(%
\begin{array}{c}
  1 \\
  0 \\
\end{array}%
\right),
\\
b(u)\left(%
\begin{array}{c}
  1 \\
  0 \\
\end{array}%
\right)&=&-\frac{\mathrm{i}c'(u)}{u+\mathrm{i}c'(u)}\left(%
\begin{array}{c}
  0 \\
  1 \\
\end{array}%
\right),
\\
c(u)\left(%
\begin{array}{c}
  1 \\
  0 \\
\end{array}%
\right) &=& 0,
\\
d(u)\left(%
\begin{array}{c}
  1 \\
  0 \\
\end{array}%
\right)&=&\frac{u}{u+\mathrm{i}c'(u)}\left(%
\begin{array}{c}
  1 \\
  0 \\
\end{array}%
\right).
\end{eqnarray}

Without any loss of generality, we define the vacuum as
\begin{equation}
|\Omega\rangle=\left(%
\begin{array}{c}
  1 \\
  0 \\
\end{array}%
\right)_{1}\otimes\left(%
\begin{array}{c}
  1 \\
  0 \\
\end{array}%
\right)_{2}\otimes\ldots\otimes\left(%
\begin{array}{c}
  1 \\
  0 \\
\end{array}%
\right)_{N}.
\end{equation}
Hence the action of the monodromy matrix on this state is
\begin{eqnarray}
\nonumber T(u)|\Omega\rangle &=& L_{1}(k_{1}-u)\left(%
\begin{array}{c}
  1 \\
  0 \\
\end{array}%
\right)_{1}\otimes\ldots\otimes L_{N}(k_{N}-u)\left(%
\begin{array}{c}
  1 \\
  0 \\
\end{array}%
\right)_{N} \\ &\equiv& \left(%
\begin{array}{cc}
  A(u) & B(u) \\
  C(u) & D(u) \\
\end{array}%
\right)|\Omega\rangle.
\end{eqnarray}
Thus the vacuum $|\Omega\rangle$ is an eigenvector  of $A(u)$,
$C(u)$ and $D(u)$ with eigenvalues $\prod_{j=1}^{N}a(k_{j}-u)$, 0
and $\prod_{j=1}^{N}d(k_{j}-u)$, respectively. Meanwhile, $B(u)$
acts as a creation operator for spin-downs.

Any arbitrary state $\Phi(\lambda)$ can be created in the form of
\begin{equation}
\Phi(\lambda)=B(\lambda_{1})B({\lambda_{2}})\ldots
B({\lambda_{M}})|\Omega\rangle,
\end{equation}
where $M$ denotes the number of spin-downs in the system. The action
of the monodromy matrix on this arbitrary state gives
\begin{equation}
T(\mu)\Phi(\lambda)=\left(%
\begin{array}{cc}
  A(\mu) & B(\mu) \\
  C(\mu) & D(\mu) \\
\end{array}%
\right)B(\lambda_{1})B({\lambda_{2}})\ldots
B({\lambda_{M}})|\Omega\rangle.
\end{equation}
Since the transfer matrix is the trace of the monodromy matrix over
the auxiliary space, we only need to consider
$A(\mu)B(\lambda_{1})B({\lambda_{2}})\ldots
B({\lambda_{M}})|\Omega\rangle$ and
$D(\mu)B(\lambda_{1})B({\lambda_{2}})\ldots
B({\lambda_{M}})|\Omega\rangle$.

From the Yang-Baxter equation of the form given in
Eq.~\eqref{YangBaxterT}, we obtain the commutation relations
\begin{eqnarray}
&&[A(u),A(v)]=0, \qquad [B(u),B(v)]=0, \\
&&[C(u),C(v)]=0, \qquad [D(u),D(v)]=0, \\
&&A(u)B(v)=\frac{u-v-\mathrm{i}c'(u-v)}{u-v}B(v)A(u)+\frac{\mathrm{i}c'(u-v)}{u-v}B(u)A(v),
\\
&&D(u)B(v)=\frac{v-u-\mathrm{i}c'(v-u)}{v-u}B(v)D(u)+\frac{\mathrm{i}c'(v-u)}{v-u}B(u)D(v),
\end{eqnarray}
where we took a negative factor in the argument of the $R$-matrix
because the arguments of the $R$-matrices in Eq.~\eqref{monodromy}
are negative with respect to $u$. Therefore
\begin{eqnarray}
\nonumber \lefteqn{A(\mu)B(\lambda_{1})B({\lambda_{2}})\ldots
B({\lambda_{M}})|\Omega\rangle}
\\ && =\prod_{i=1}^{M}\frac{\mu-\lambda_{i}-\mathrm{i}c'(\mu-\lambda_{i})}{\mu-\lambda_{i}}
\prod_{l=1}^{N}\frac{\mu-k_{l}+\mathrm{i}c'(\mu-k_{l})}{\mu-k_{l}-\mathrm{i}c'(\mu-k_{l})}|\Omega\rangle+\textrm{unwanted
terms},\label{Au}
\end{eqnarray}
and
\begin{eqnarray}
\nonumber \lefteqn{D(\mu)B(\lambda_{1})B({\lambda_{2}})\ldots
B({\lambda_{M}})|\Omega\rangle}
\\ && =\prod_{i=1}^{M}\frac{\mu-\lambda_{i}+\mathrm{i}c'(\mu-\lambda_{i})}{\mu-\lambda_{i}}
\prod_{l=1}^{N}\frac{\mu-k_{l}}{\mu-k_{l}-\mathrm{i}c'(\mu-k_{l})}|\Omega\rangle+\textrm{unwanted
terms.}\label{Du}
\end{eqnarray}
The sum of the unwanted terms in Eqs.~\eqref{Au} and \eqref{Du}
vanish when there are no poles in the eigenvalue of
Eq.~\eqref{eigenEquation2}.

From Eq.~\eqref{eigenEquation2}, we obtain the ABA equations
\begin{equation}
\exp(\mathrm{i}k_{j}L)=
-\prod_{l=1}^{N}\frac{k_{j}-k_{l}+\mathrm{i}c'(k_{j}-k_{l})}{k_{j}-k_{l}-\mathrm{i}c'(k_{j}-k_{l})}
\prod_{i=1}^{M}\frac{k_{j}-\lambda_{i}-\mathrm{i}c'(k_{j}-\lambda_{i})}{k_{j}-\lambda_{i}},
\qquad j=1,\ldots,N,
\end{equation}
\begin{equation}
\prod_{l=1}^{N}\frac{\lambda_{i}-k_{l}+\mathrm{i}c'(\lambda_{i}-k_{l})}{\lambda_{i}-k_{l}}
=-\prod_{j=1}^{M}\frac{\lambda_{i}-\lambda_{j}+\mathrm{i}c'(\lambda_{i}-\lambda_{j})}
{\lambda_{i}-\lambda_{j}-\mathrm{i}c'(\lambda_{i}-\lambda_{j})},
\qquad i=1,\ldots,M.
\end{equation}
Note that here we cannot make a uniform shift for the set
$\{\lambda_{i}\}$, i.e.,
$\lambda_{i}\rightarrow\lambda_{i}-\mathrm{i}c/2$ for every $i$,
because the effective interaction strength $c'(u)$ depends on the
quasimomenta $\{k_{j}\}$ and the rapidities $\{\lambda_{i}\}$.

\end{document}